\documentclass[a4paper,fleqn,usenatbib]{mnras}

\usepackage[T1]{fontenc}
\usepackage{ae,aecompl}

\usepackage{graphicx}	
\usepackage{amsmath}	
\usepackage{amssymb}	
\usepackage{nccmath}
\usepackage{siunitx}
\usepackage{natbib}
\usepackage{subfigure}
\usepackage[usenames]{color}

\title[High-energy processes in starburst-driven winds]{High-energy processes in starburst-driven winds}

\author[M\"{u}ller, Romero \& Roth]{
Ana L. M\"{u}ller$^{1,3,4}$\thanks{E-mail: almuller@iar-conicet.gov.ar},
Gustavo E. Romero$^{1,2}$
and Markus Roth$^{3}$\\
$^{1}$ Instituto Argentino de Radioastronom\'{\i}a (CONICET; CICPBA), C.C. No. 5, 1894 Villa Elisa, Argentina.\\
$^{2}$ Facultad de Ciencias Astron\'omicas y Geof\'{\i}sicas, Universidad Nacional de La Plata, Paseo del Bosque s/n, 1900, La Plata, Argentina.\\
$^{3}$ Institute for Nuclear Physics (IKP), Karlsruhe Institute of Technology (KIT), Germany. \\
$^{4}$ Instituto de Tecnolog\'{\i}as en Detecci\'on y Astropart\'{\i}culas (CNEA, CONICET, UNSAM), Buenos Aires, Argentina.}

\pubyear{2019}
\graphicspath{./figs/}

\begin{document}
\label{firstpage}
\pagerange{\pageref{firstpage}--\pageref{lastpage}}
\maketitle

\begin{abstract}
Starburst galaxies generate large-scale winds powered by the activity in the star-forming regions located in the galactic disks. Fragmentation of the disk produced by the  outbreak of the wind results in the formation of clouds. Bowshocks caused by the supersonic outflow appear around such clouds. In this paper we discuss the acceleration of relativistic particles and the production of non-thermal radiation in such scenario. Cosmic rays accelerated at the bowshocks do not reach the highest energies, although the high-energy luminosity generated is significant. We show that up to $\sim10$\% of the gamma-ray emission in starbursts might come from these sources outside the galactic disks. Discrete X-ray sources with a power-law component are also expected. 
\end{abstract}




\section{Introduction}

Starburst galaxies have intense episodes of star formation in their galactic disks. This activity results in the formation of a galactic wind that breaks out from the disk and expands into the halo of the galaxy, sweeping gas and forming a hot region that is usually detected in X-rays. The galactic wind transports metals created in the disk and injects them into the halo and the intergalactic medium (for a recent review see, e.g., \cite{veilleux2005}). The standard model for the production of galactic winds was proposed long ago by \cite{chevalier1985}: the combined effect of supernova explosions and stellar winds creates a very hot bubble in the star forming region \mbox{($T\sim 10^8$ K)}. The internal pressure of this gas is so high that it exceeds the gravitational binding energy and the gas disrupts the disk, expanding adiabatically through the halo and dragging with it fragments of the cold matter that formed the disk. The wind sweeps the ambient gas creating a multi-phased bubble with cold, warm, and hot components \citep{strickland2002}.

Because of the existence of multiple shocks, a high-metallicity environment, and a huge energy budget, starbursts are considered as sites of non-thermal particle acceleration and high-energy radiation \citep{paglione1996,bykov2001,romero2003b,domingo2005,rephaeli2010,bykov2014,peretti2018}. This has been confirmed by the gamma-ray detection of nearby starburst galaxies \citep{acero2009,abdo2010,ackermann2012,ohm2016}. 

The indication of a non-zero metallicity content in the ultra high-energy cosmic ray spectrum also suggests nearby starbursts as possible sites of cosmic ray acceleration up to energy of around \mbox{$10^{20}$ eV}. This was first proposed by  \cite{anchordoqui1999} and recently revisited by \cite{anchordoqui2018} and \cite{romero2018}. However, \cite{romero2018} have found that the conditions necessary to achieve energies of \mbox{$\sim10^{20}$ eV} in the hot wind region seem to be unphysical and at odds with the observational data. Typical velocities of the galactic winds are of the order of 
\begin{equation}
  v_{{\rm w} \infty} \approx \sqrt{2\dot{E}/\dot{M}}\sim10^3\;\;\textrm{km\; s}^{-1}, \label{vinfty} 
\end{equation}
where $\dot{E}$ and $\dot{M}$ are the total energy released in the starburst region and the mass input, respectively. The magnetic field in the halo of the galaxy NGC 253, a southern well-known galaxy with star forming activity, has been determined through radio polarization observations by \cite{heesen2009b} and is of the order of \mbox{5 $\mu$G}. The average particle density in the galactic wind bubble of radius \mbox{$R_{\rm b}\sim5$ kpc} is \mbox{$n_{\rm w}\sim 2\times10^{-3}$ cm$^{-3}$} \citep{strickland2002}. With such parameters, diffusive shock acceleration yields maximum energies of \mbox{$\sim10^{16}$} and \mbox{$\sim5\times 10^{17}$ eV} for protons and iron nuclei, respectively (see \cite{romero2018} for a detailed discussion). \cite{anchordoqui2018} invokes higher values of the magnetic field, of \mbox{$\sim300\,\mu$G}. With such a value the magnetic energy density \mbox{$u_{\rm B}=B^{2}/8\pi$} is \mbox{$\sim4\times10^{-9}$ erg cm$^{-3}$}.  But the ram pressure of the gas is \mbox{$u_{\rm g}\approx n_{\rm w}\,m_p\,v_{\rm w}^2\sim 10^{-11}$ erg cm$^{-3}$}, so the magnetization parameter results:
\begin{equation}
  \beta=\frac{u_{\rm B}}{u_{\rm g}}>>1.  \label{mag}
\end{equation}
Therefore the medium is mechanically incompressible and the shock cannot exist under such conditions. 

One way around this situation is to invoke magnetic field amplification in the presence of shocks. This is known to operate in galactic supernova remnants \citep{bamba2003,vink2003}. It has been suggested that the mechanism responsible for this amplification is the non-resonant hybrid (NRH) instability (also known as the Bell instability, \cite{bell2004}). In the non-linear regime this effect can produce an amplification of up to two orders of magnitude of the field in regions of originally low magnetization \citep{matthews2017}. If there are high-density regions in the wind, in such a way that the ram pressure of the gas be several orders of magnitude larger than the average, then amplification from the initial few \mbox{$\mu$G} field to values close to \mbox{$1$ mG} might occur.  A natural site to explore this possibility is in the bowshocks formed around dense inhomogeneities in the halo. 

In this paper we study particle acceleration in the bowshocks formed by the galactic wind of a standard starburst around clouds in the halo. These clouds are fragments of the disk that are dragged by the outflow. In the next section we discuss the scenario in more detail. Some preliminary results were shown by \citet{HEPRO2019}. Here we shall show that although ultra high-energies are not reached, these bowshocks inject a considerable amount of cosmic rays up to \mbox{$\sim 10^{17}$ eV} and produce X-ray and $\gamma$-emission that could be detectable. In Section  \ref{sec:cloud-wind} we present the physics of the wind-cloud interaction and estimate different relevant timescales. Section \ref{sec:acc} is devoted to the estimate of the particle acceleration, different losses, and the resulting particle distributions. The radiation produced by these non-thermal particles is computed and shown in Section \ref{sec:rad}. Section \ref{sec:discu} presents a discussion of our results and their implications. We close with a summary and some brief conclusions in Section \ref{sec:concl}.

\section{Origin of the clouds in the halo of starbursts}\label{sec:clouds}

\begin{figure}
\centering
\includegraphics[width=0.45\textwidth,keepaspectratio]{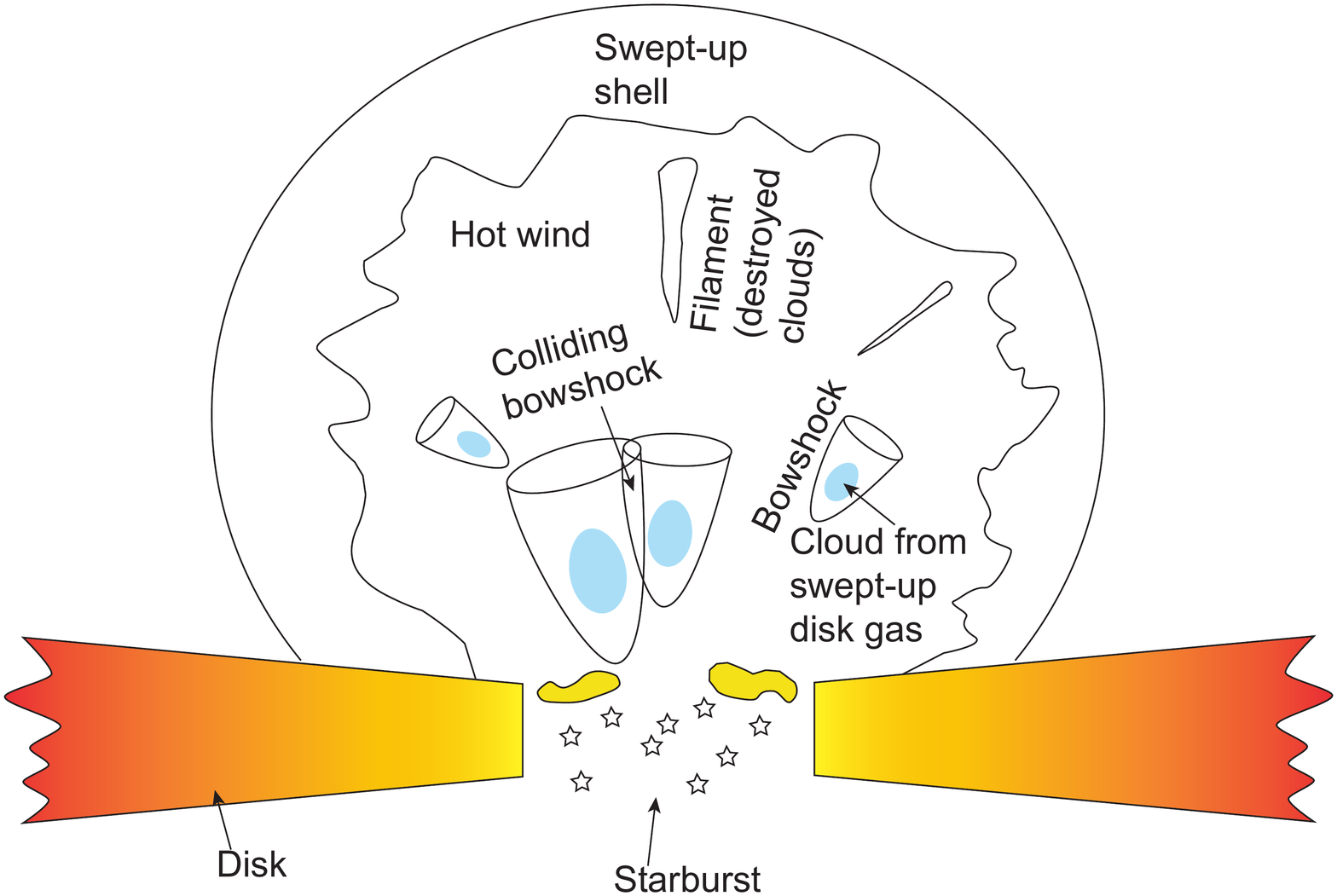}
\caption{Scheme of starbursts hot wind interacting with several clouds in the halo (not to scale). Based on \protect\cite{cooper2008}.}
\label{fig:wind}
\end{figure}

The development of a galactic wind powered by a central region with a high star formation rate has been modeled both analytically and through numerical simulations \citep{chevalier1985,strickland2000,cooper2008}. The wind originates through many different hot bubbles in the central starburst. These bubbles expand and merge leading to the formation of a large, very hot cavity that disrupts the disk after \mbox{$\sim 0.15$ Myr} \citep{cooper2008}. The distribution of gas in the disk is not homogeneous, so the disk undergoes fragmentation and clumps are dragged with the outflow. The wind quickly reaches velocities of \mbox{$\sim 10^3$ km s$^{-1}$}, exerting pressure onto the clouds and fragments, which are accelerated along the flow lines. Since the velocity of the flow is highly supersonic, bowshocks are formed around the different clouds. The overall picture is depicted in Fig. \ref{fig:wind}.

The 3D simulations by \cite{cooper2008} show how as time goes by clouds are ablated by the wind and the gas forms filamentary-like structures consisting of a stream of colder material  (see their Fig. 13, in particular the right panel, which corresponds to an evolution of 2 Myr). 

The acceleration of a cloud by the wind is, roughly, 
\begin{equation}
  a_{\rm ac} \approx \xi \left( \frac{n_{\rm w}}{n_{\rm c}}\right) \frac{v_{\rm w}^2}{R_{\rm c}}.
\end{equation}
Here $\xi\sim1$ is the dragging coefficient, $R_{\rm c}$ is the radius of the cloud, and $n_{\rm c}$ is the density of the cloud. A cloud of \mbox{$R_{\rm c}=5$ pc} and average density of \mbox{$n_{\rm c}=100$ cm$^{-3}$}, immersed in a wind with \mbox{$v_{\rm w}=2000$ km s$^{-1}$} and average number density \mbox{$n_{\rm w}=10^{-3}$ cm$^{-3}$}, would have an acceleration \mbox{$a_{\rm ac}\sim 10^{-13}$ km s$^{-2}$}. Such acceleration indicates that, in general, clouds will only reach modest velocities during the existence of the starburst episode (a few Myr). 

This estimate does not take into account the details of the hydrodynamics, the ablation of the cloud, the effects of shocks, etc. But the result shows that, in general, the wind will move at highly supersonic velocity with respect to the cloud and bowshocks will be formed around clouds at different stages of the evolution of the starburst.  

\section{Cloud-wind interactions in starbursts}\label{sec:cloud-wind}

\begin{figure}
\centering
\begin{center}
\includegraphics[scale=.4,trim=5cm 0cm 0cm 0cm, clip=true,angle=0]{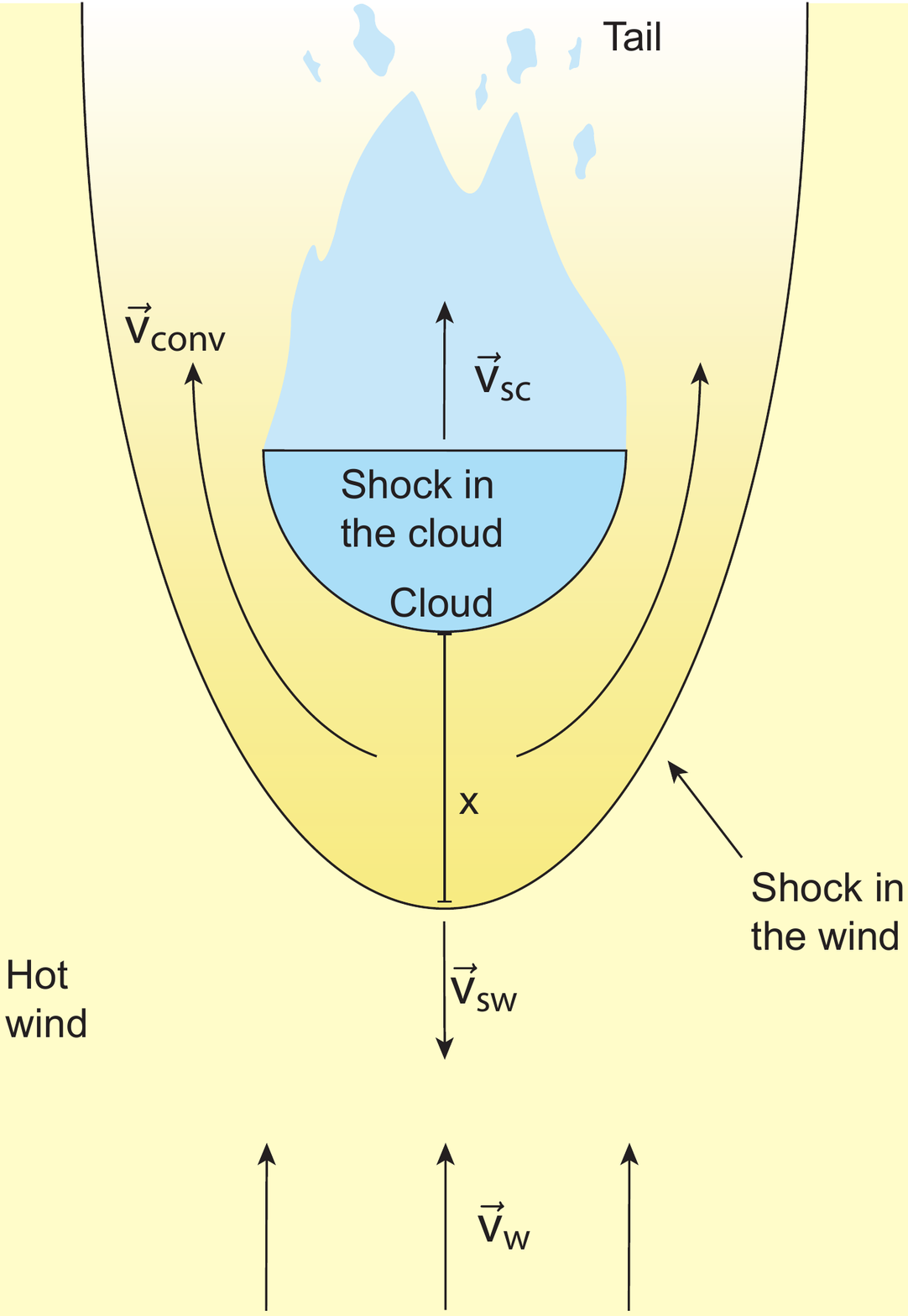}
\caption{Schematic diagram of the bowshock around a cloud immersed in the starbursts hot wind (not to scale).}
\label{fig:cloud}
\end{center}
\end{figure}

The interaction of a cloud with a hot wind has been investigated by many authors. \cite{klein1994} identify four evolutionary phases in the interaction. First, when the cloud is reached by the wind, a system of two shocks is formed: one shock moves through the cloud and the other propagates backwards through the wind.  
A bowshock then appears around the cloud, with a contact discontinuity located at a minimum distance (at the bowshock apex) of $x\sim 0.2 R_{\rm c}$ \citep{vanDyke1959}. The compressed gas of the wind flows inside this region with a velocity $v_{\rm conv}$ triggering Kelvin-Helmholtz (KH) instabilities. Frontal pressure on the cloud can result in Rayleigh-Taylor (RT) instabilities. The impact of the shock on the rear part of the cloud produces a turbulent rarefaction that forms a tail of gas. In the last phase the cloud fragments and is destroyed by the instabilities. The different elements described are schematically represented in \mbox{Fig. \ref{fig:cloud}}.

Different timescales are associated with the cloud-wind interaction. The cloud-crushing time is defined as \citep{klein1994}:

\begin{equation}
  t_{\rm crush}\approx \frac{R_{\rm c}}{v_{\rm sc}}\approx \frac{R_{\rm c}}{v_{\rm w}} \sqrt{\frac{n_{\rm c}}{n_{\rm w}}},
\end{equation}
where $v_{sc}$ is the velocity of the shock-in-the-cloud.

Then, we have the timescales for the KH and RT instabilities, given by:

\begin{equation}
  t_{\rm KH}=\frac{R_{\rm c}(n_{\rm c} + n_{\rm w} )}{(v_{\rm w}-v_{\rm c}) (n_{\rm c}  n_{\rm w})^{1/2} },\end{equation}
and
\begin{equation}
  t_{\rm RT}=\sqrt{\frac{R_{\rm c}}{a_{\rm c}}}\approx \frac{R_{\rm c}}{v_{\rm w}} \sqrt{\frac{n_{\rm c}}{n_{\rm w}}}.
\end{equation}
In these expressions we have considered the instability length as of the order of the radius of the cloud. 

Numerical simulations in 2 and 3D clearly show that the ablation process of the cloud is strongly dependent on the morphology and density of the cloud, the efficiency of radiative cooling, and the presence of magnetic fields \citep{xu1995,gregori2000,shin2008,cooper2009,mccourt2015,sparre2019}. Fractal clouds disrupt much faster than spherical ones \citep{cooper2009}. The presence of a significant magnetic field tends to make the cloud more resistant and cohesive \citep{mccourt2015}. Radiative cooling plays also an important role in the survival of the cloud. An adiabatic cloud inflates as a consequence of the injection of energy. This increases its radius and aggravates the effects of the different instabilities. In an adiabatic cloud the material starts to be ablated almost immediately; instead when all the injected heat is radiated away, the ablation is significantly delayed \citep{cooper2009}. A radiative cloud also remains cold, conversely to adiabatic ones which reach temperatures up to \mbox{$10^6$ K}.

 The thermal cooling time for the cloud is:
 \begin{equation}
   t_{\rm \Lambda}=\frac{R_{\rm \Lambda}}{v_{\rm sc}},
 \end{equation}  
where $R_{\rm \Lambda}$ is the cooling length given by \citep{mccray1979}:
 \begin{equation}
    R_{\Lambda}=\frac{1.90\times10^{-29}\, \mu\, (v_{\textrm{sc}}/\textrm{km s$^{-1}$})^{3}}{(n/ \textrm{cm$^{-3}$})\,(\Lambda(T)/\textrm{erg cm$^{3}$ s$^{-1}$})}\textrm{ pc}
   \label{eqn:thermallength}
   \end{equation}
   
   \begin{equation}
    \textrm{ with } \,\,\, T=18.21\, \mu \, \left( \frac{v_{\textrm{sc}}}{\textrm{km s$^{-1}$}} \right)^{2} \text{ K}.
   \end{equation}

  \noindent Here, $\mu$ is 0.6 if the material is ionized or 1.3 if it is neutral, and $\Lambda(T)$ [erg cm$^{3}$ s$^{-1}$] is the cooling function \citep{wolfire2003,raymond1976,myasnikov1998}:
  
  \begin{equation}
   \Lambda(T)=\left\{
	       \begin{array}{llll}
	     4\times 10^{-29}T^{0.8} & \mathrm{if\ } 55\,{\rm K} \le T < 10^{4}\,{\rm K}\\ 
		 7\times 10^{-27}T  & \mathrm{if\ } 10^{4}\,{\rm K} \le T < 10^{5}\,{\rm K} \\
		 7\times 10^{-19}T^{-0.6} & \mathrm{if\ } 10^{5}\,{\rm K} \le T < 4\times 10^{7}\,{\rm K} \\
		 3\times 10^{-27}T^{0.5} & \mathrm{if\ } T \ge 4\times 10^{7}\,{\rm K}  \\
	       \end{array}
	     \right.
  \end{equation}
  
  The relation between $v_{\rm w}$, $v_{\rm sw}$ and $v_{\rm sc}$ is given by \citep[see, e.g.,][]{tenorio-tagle1981}:

\begin{equation}\label{eq:vsc}
v_{\rm sc} = - \frac{4}{3} \;\frac{1}{1 + \sqrt{n_{\rm c}/n_{\rm w}}}\,v_{\rm w},
\end{equation}

\begin{equation}\label{eq:vsd}
v_{\rm sw} =  \frac{4}{3} \;\frac{1}{1 + \sqrt{n_{\rm w}/n_{\rm c}}}\,v_{\rm w}.
\end{equation}

\noindent In these expressions we adopt a polytropic index  \mbox{$\gamma_{\rm gas} = 5/3$} for monatomic gases.

Since the density contrast between wind and cloud is of several orders of magnitude, the shock in the wind will be fast and adiabatic, and a suitable site for diffusive shock acceleration (DSA) of charged particles. Whether the shock in the cloud is capable of such acceleration will depend on the radiative cooling: strong radiative cooling significantly increases the entropy of the gas and destroys the inhomogeneities in the magnetic field that scatter the particles across the shock. 

In order to make quantitative estimates, we will calculate two different models. Model M1 consists of a small cloud of spherical shape with radius \mbox{$R_{\rm c}=5$ pc} and density \mbox{$n_{\rm c}=10^2$ cm$^{-3}$}. Model M2  is made up of a larger cloud with \mbox{$R_{\rm c}=100$ pc} and average density \mbox{$n_{\rm c}=10$ cm$^{-3}$}. These densities are typical of the disk of an average starburst. We have chosen these values since they match those adopted in the simulations by \cite{cooper2009} (M1) and \cite{sparre2019} (M2)\footnote{Both works cited are 3D HD sets of simulations of clouds being ablated by winds in a starburst context. The \cite{sparre2019}'s  set is a state-of-the-art research that includes comparisons with previous simulations, including those of \cite{cooper2009}. The findings of this earlier work are confirmed by the newer, higher resolution, simulations. Hence it is appropriate to adopt both sets to characterize the HD of our two fiducial cloud models.}. They are typical of a small and a large cloud produced in the disk fragmentation. In both cases the wind velocity is taken as \mbox{$v_{\rm w}=1000$ km s$^{-1}$} and its number density \mbox{$n_{\rm w}=10^{-2}$ cm$^{-3}$}, according to the mentioned simulations. Regarding the magnetization, the field is fixed in such a way that the magnetization parameter is \mbox{$\beta=0.9$}, both in the shocked cloud and in the shocked wind, as expected if some efficient magnetic amplification mechanism operates through instabilities in the shocked gas. This avoids the kind of problems presented by the treatment performed by \cite{anchordoqui2018} (because we have $\beta<1$), but still allows for significant magnetic field amplification from the value of a \mbox{few $\mu$G} expected in the unshocked hot wind. We summarize the parameters of the models in Table  \ref{table:one}.

\begin{table}
\caption{Parameters of the models. The magnetization \mbox{$\beta=0.9$} and the wind velocity \mbox{$v_{\rm w}=1000$ km s$^{-1}$} are the same in both cases.}
\begin{center}
\begin{tabular}{llllll}
\hline
Model &  $R_{\rm c}$  &$n_{\rm w}$ & $n_{\rm c}$ &  $v_{\rm sw}$  & $v_{\rm sc}$ \\
& [pc] &  [cm$^{-3}$] & [cm$^{-3}$]  &   [km~s$^{-1}$] & [km~s$^{-1}$] \\
\hline
M1 & 5 & $0.01$ & $100$ & 1320 & 4.2  \\
M2 & 100 & 0.01 & 10 & 1292 & 13.2 \\
\hline
\end{tabular}
\end{center}
\label{table:one}
\end{table}

The small shock velocities in the cloud clearly indicate that the shocks are radiative and inefficient for particle acceleration. DSA only can occur in the reverse shocks in the wind. The magnetic field in the acceleration region is:

\begin{ceqn}
\begin{equation}
  u_{\rm B}=0.9\,u_{\rm g}
  \nonumber
\end{equation}
\begin{equation}
\centering
  \frac{B^{2}_{\rm sw}}{8\,\pi}=0.9\,\frac{3}{2}P_{\rm ram}=0.9\,\frac{9}{8}\,n_{w}\,m_{p}\,v^{2}_{\rm sw}
  \nonumber
\end{equation}
\begin{equation}
\centering
  B_{\rm sw}=\sqrt{\frac{81}{10}\,\pi\,n_{w}\,m_{p}}\,v_{\rm sw}.
\end{equation}
\end{ceqn}

\noindent This yields a magnetic field of \mbox{$B_{\rm sw}\sim 86\, \mu$G} for M1 and \mbox{$B_{\rm sw}\sim 84\, \mu$G} for M2. 

In what follows we investigate the acceleration and radiation of particles in the reverse shock in the galactic wind around the clouds characterized by models M1 and M2. The different dynamical timescales are presented in Table \ref{table:2}. It can be seen that the shocks in the cloud are fully radiative and that the dynamical lifetime of the cloud is set by the cloud-crushing time for both models.

\begin{table}
\caption{Dynamical timescales calculated according to the formulas given in the text.}
\begin{center}
\begin{tabular}{llllll}
\hline
Model &  $t_{\rm crush}$  &$t_{\rm KH}$ & $t_{\rm RT}$ &  $t_{{\rm \Lambda}_{\rm sc}}$ & $t_{{\rm \Lambda}_{\rm sw}}$  \\
& [Myr] &  [Myr] & [Myr]  &   [Myr] &   [Myr] \\
\hline
M1 & $0.37$ & $0.49$ & $0.37$ & $1.15\times10^{-3}$ & $64.78$   \\
M2 & $2.39$ & $3.09$ & $2.39$ & $1.46\times10^{-5}$ & $60.53$  \\
\hline
\end{tabular}
\end{center}
\label{table:2}
\end{table}

\section{Particle acceleration, losses, and diffusion}\label{sec:acc}

Charged particles can be accelerated in the reverse shock in the wind by DSA. This kind of situation has been studied by several authors, although never in the present context \citep[see, e.g.,][]{araudo2009,araudo2010,delvalle2012,delvalle2018a,delvalle2018b,delpalacio2018}.

The acceleration region has a size $x$ (see Fig. \ref{fig:wind}). The Hillas criterion imposes an absolute upper limit to the energy that particles can achieve:
\begin{equation}
  E_{\rm max}=10^{15}Z\left(\frac{x}{\rm pc}\right)\left(\frac{B}{\mu\rm G}\right)\,{\rm eV}  \label{Hillas},
\end{equation}
where $Z$ is the atomic charge number. We obtain the following results for protons and iron nuclei in the models considered here (Table \ref{table:one}):

Model M1:
 \begin{eqnarray}
  E^p_{\textrm{max}}&=&8.6\times10^{16}\,\textrm{eV} \;\;\;\;\textrm{protons} \label{Epmax1M1}\\
  E^{\textrm{Fe}}_{\textrm{max}}&=&2.2\times10^{18}\,\textrm{eV} \;\;\;\;\textrm{iron nuclei} \label{Efemax1M1}
  \end{eqnarray}

Model M2:
\begin{eqnarray}
  E^p_{\textrm{max}}&=&1.7\times10^{18}\,\textrm{eV} \;\;\;\;\textrm{protons} \label{Epmax1M2}\\
  E^{\textrm{Fe}}_{\textrm{max}}&=&4.4\times10^{19}\,\textrm{eV} \;\;\;\;\textrm{iron nuclei} \label{Efemax1M2}
\end{eqnarray}

Radiative and spatial losses will further restrict the maximum energy of the particles. Protons, in this scenario, will be affected by convection from the acceleration region by the wind. The timescale of this process is \mbox{$t_{\rm conv}\approx 4R_{\rm c}/v_{\rm sw}$}. Then,
\begin{eqnarray}
  t^{\rm M1}_{\rm conv}&\sim&4.7\times10^{11}\,{\rm s}\,\sim 1.5\times10^{-2}\,{\rm Myr}   \label{tconv1}\\
   t^{\rm M2}_{\rm conv}&\sim&9.5\times10^{12}\,{\rm s}\,\sim 0.30 \,{\rm Myr} \label{tconv2}  
  \end{eqnarray}

Diffusion of protons both upstream and downstream can also be important. For the acceleration the Bohm diffusion is a good approximation. The diffusion timescale in Bohm's regimen is:

\begin{equation}
  t_{\rm diff,\, Bohm}\sim 10^{13}\left( \frac{R_{\rm c}}{\rm pc} \right)^{2}\left( \frac{B}{\mu \rm G} \right)\left( \frac{E}{\rm GeV} \right)^{-1}\,{\rm s}.\label{Bohm}
  \end{equation}
Then,
\begin{eqnarray}
  t^{\rm M1}_{\rm diff,\, Bohm}&\sim& 6.8\times10^{2} \left(\frac{E}{\rm GeV}\right)^{-1}\,{\rm Myr}  \label{tdiff1B} \\
   t^{\rm M2}_{\rm diff,\, Bohm}&\sim& 2.7\times 10^{5} \left(\frac{E}{\rm GeV}\right)^{-1}\,{\rm Myr} \label{tdiff2B}  
\end{eqnarray}

The acceleration rate by DSA in the test particle limit is given by:
\begin{equation}
  \frac{dE}{dt}=\frac{3}{20}\,e\,c\,Z\,\left( \frac{D}{D_{\textrm{B}}}\right)^{-1} \left( \frac{v_{\textrm{sw}}}{c}\right)^{2} B,
\end{equation}
where $D$ is the diffusion coefficient in the shocked wind region in Bohm units: $D_{\textrm{B}}=c\,r_{\textrm{L}}/3$. The acceleration timescale is \citep{romero2018}:

\begin{equation}
\begin{split}
  t_{\textrm{acc}}\approx 2.1\;&Z^{-1}\left( \frac{D}{D_{\textrm{B}}}\right) \\
  &\times \left( \frac{v_{\textrm{sw}}}{1000\;\textrm{km}\;\textrm{s}^{-1}} \right)^{-2} \left( \frac{B}{\mu\textrm{G}}\right)^{-1} \left(\frac{E}{\textrm{GeV}}\right)\,\textrm{yr}.
\end{split} 
\end{equation} 
With the assumed amplified magnetic field of \mbox{$86\,\mu$G} for M1 and \mbox{$84\,\mu$G} for M2 this becomes:
\begin{equation}
  t^{\rm M1}_{\textrm{acc}}\approx 2.46\times10^{-8}\, Z^{-1} \,\left(\frac{E}{\textrm{GeV}}\right)\,\textrm{Myr}
  \label{tacc1B}
\end{equation}
\begin{equation}
  t^{\rm M2}_{\textrm{acc}}\approx 2.52\times10^{-8}\, Z^{-1} \,\left(\frac{E}{\textrm{GeV}}\right)\,\textrm{Myr}.
  \label{tacc2B}
\end{equation}

\begin{figure*} 
\subfigure{\includegraphics[trim= 0cm 0cm 0cm 0cm, clip=true, width=.47\textwidth,angle=0]{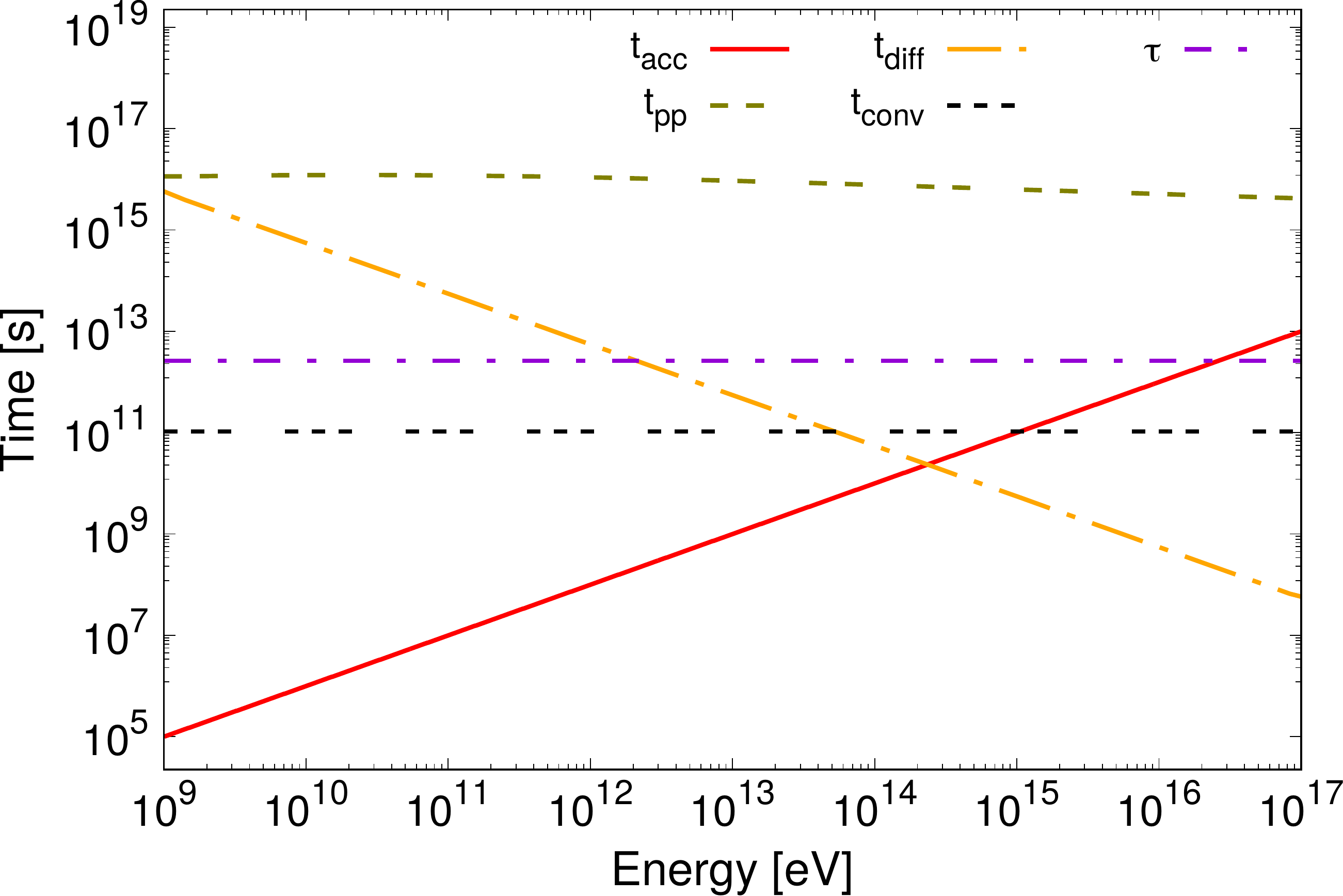}}
\subfigure{\includegraphics[trim= 0cm 0cm 0cm 0cm, clip=true, width=.47\textwidth,angle=0]{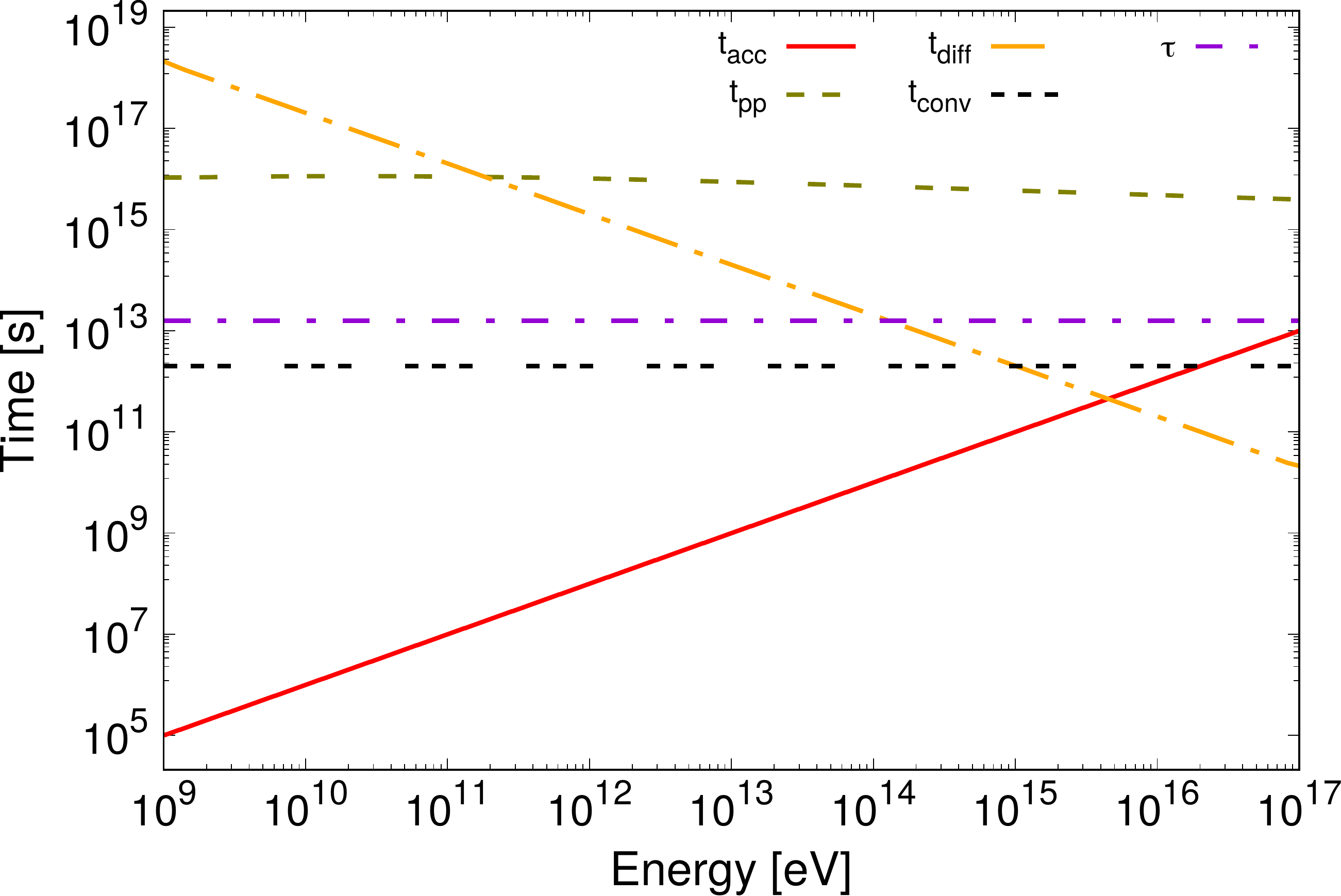}}
\caption{Acceleration and cooling timescales for the protons in both models. On the left we show the case of the small cloud M1, and on the right the results for the massive M2 cloud. $\tau$ is the dominant dynamical timescale of the systems.} 
\label{fig:coolingprotons}
\end{figure*}

Radiative losses for protons are negligible during the acceleration, so their maximum energy will be determined by the removal of particles caused by diffusion (see \mbox{Fig. \ref{fig:coolingprotons}}). Then, matching Eqs. (\ref{tacc1B}) and (\ref{tacc2B}) with Eqs. (\ref{tdiff1B}) and (\ref{tdiff2B}), we find:

Model M1:
 \begin{eqnarray}
  E^p_{\textrm{max}}&=&1.7\times10^{14}\,\textrm{eV} \;\;\;\;\textrm{protons} \label{EpmaxM1}\\
  E^{\textrm{Fe}}_{\textrm{max}}&=&4.3\times10^{15}\,\textrm{eV} \;\;\;\;\textrm{iron nuclei} \label{EfemaxM2}
  \end{eqnarray}

Model M2:
\begin{eqnarray}
  E^p_{\textrm{max}}&=&3.3\times10^{15}\,\textrm{eV} \;\;\;\;\textrm{protons} \label{Epmax2M1}\\
  E^{\textrm{Fe}}_{\textrm{max}}&=&8.6\times10^{16}\,\textrm{eV} \;\;\;\;\textrm{iron nuclei} \label{Efemax2M1}
\end{eqnarray}
These values are much more modest than the maximum ones allowed by Hillas criterion (Eqs.(\ref{Epmax1M1}) -- (\ref{Efemax1M2})). They show that ultra-high energy cosmic rays cannot be produced in the scenario discussed here.

\begin{figure*} 
\hspace{-0.12cm}\subfigure{\includegraphics[trim= 0cm 0cm 0cm 0cm, clip=true, width=.459\textwidth,angle=0]{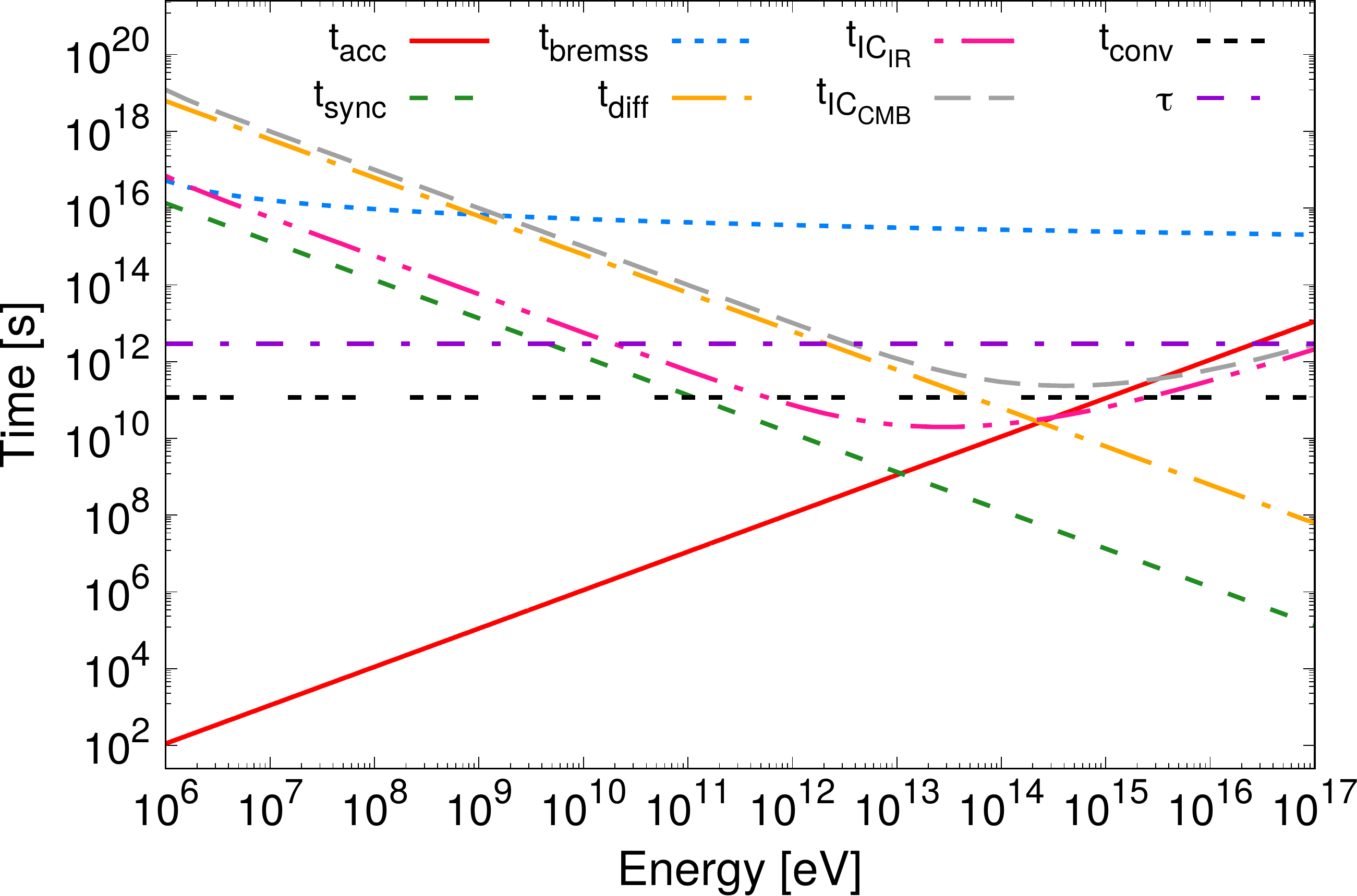}\label{fig:coolingelectronsM1}}\hspace{0.18cm}
\subfigure{\includegraphics[trim= 0cm 0cm 0cm 0cm, clip=true, width=.459\textwidth,angle=0]{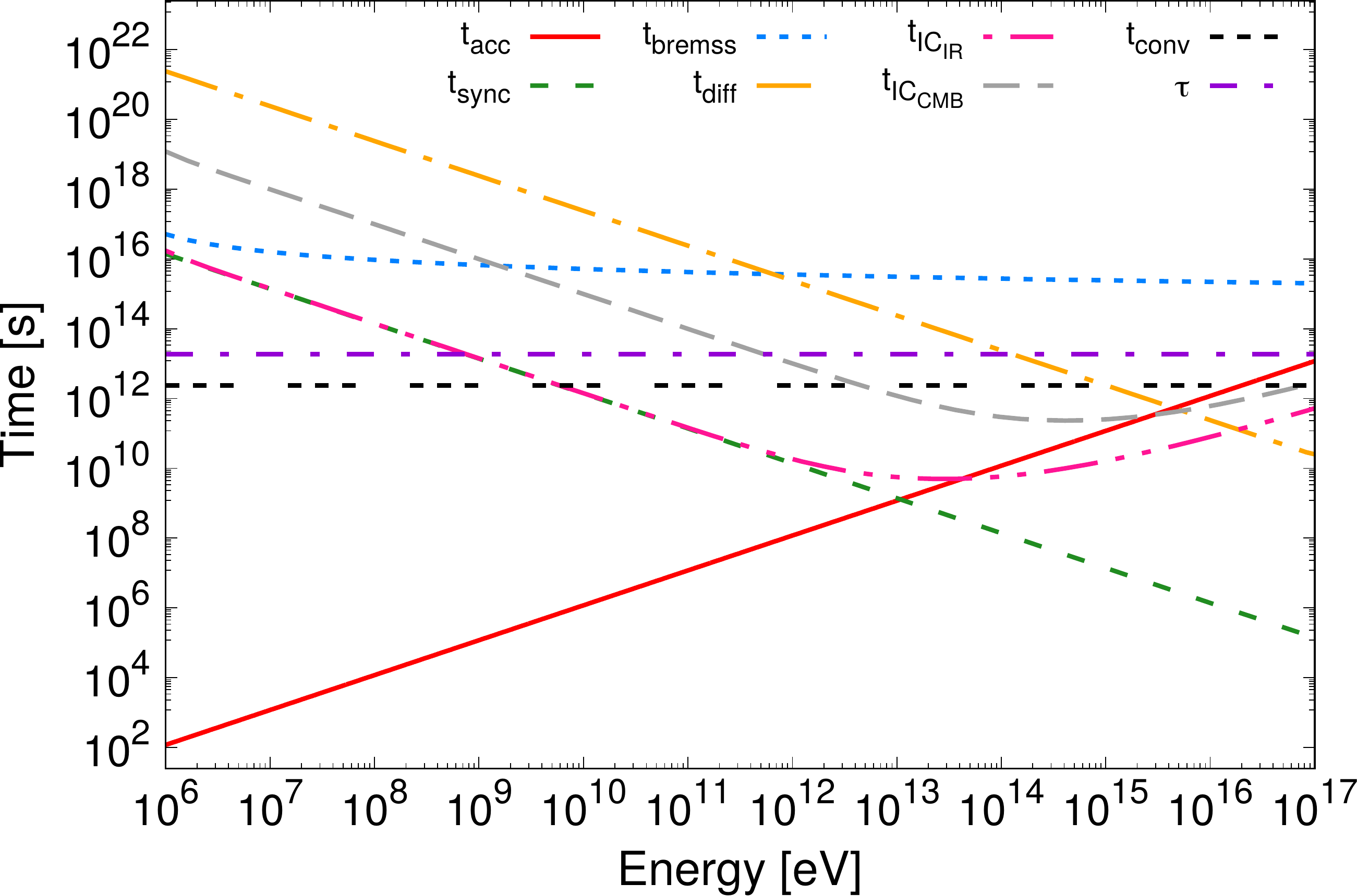}\label{fig:coolingelectronsM2}}
\caption{Acceleration and cooling timescales for the electrons in both models. On the left we show the case of the small cloud M1, and on the right the results for the massive M2 cloud. $\tau$ is the dominant dynamical timescale of the systems.} 
\label{fig:coolingelectrons}
\end{figure*}

The radiative losses for electrons include synchrotron radiation, relativistic Bremsstrahlung, and inverse Compton scattering of CMB and IR photons. In the case of the IR emission, we assume its luminosity has a typical value of \mbox{$10^{10.5}$ L$_{\odot}$} and it is produced by a blackbody whose temperature is \mbox{$40$ K}. The energy density of the IR radiation field decreases with the square of the distance from the galactic plane \citep{lacki2013}. Since larger clouds are expected to exist closer to the disk, we adopt a distance of \mbox{$500$ pc} for M2 and \mbox{$1$ kpc} for M1. Expressions for the calculation of these losses in the present setting are given by \cite{romero2018}. The maximum energy for electrons will be defined in M1 by the synchrotron emission (see Fig. \ref{fig:coolingelectrons}, left panel). The cooling timescales for synchrotron and IC with the IR photons are similar for M2 (see Fig. \ref{fig:coolingelectrons}, right panel), thus the maximum energy is given by \mbox{$t^{-1}_{\rm acc}\approx t^{-1}_{\rm synchr}+t^{-1}_{\rm IC}\approx 2\,t^{-1}_{\rm synchr}$}. Then, the values obtained are \mbox{$E^{e}_{\rm max}=6.7\times10^{12}$ eV} for M1 and \mbox{$E^{e}_{\rm max}=4.9\times10^{12}$ eV} for M2.

\section{Radiation}\label{sec:rad}

\begin{figure*}
\centering
\subfigure{\includegraphics[trim= 0cm 0cm 0cm 0cm, clip=true, width=.46\textwidth,angle=0]{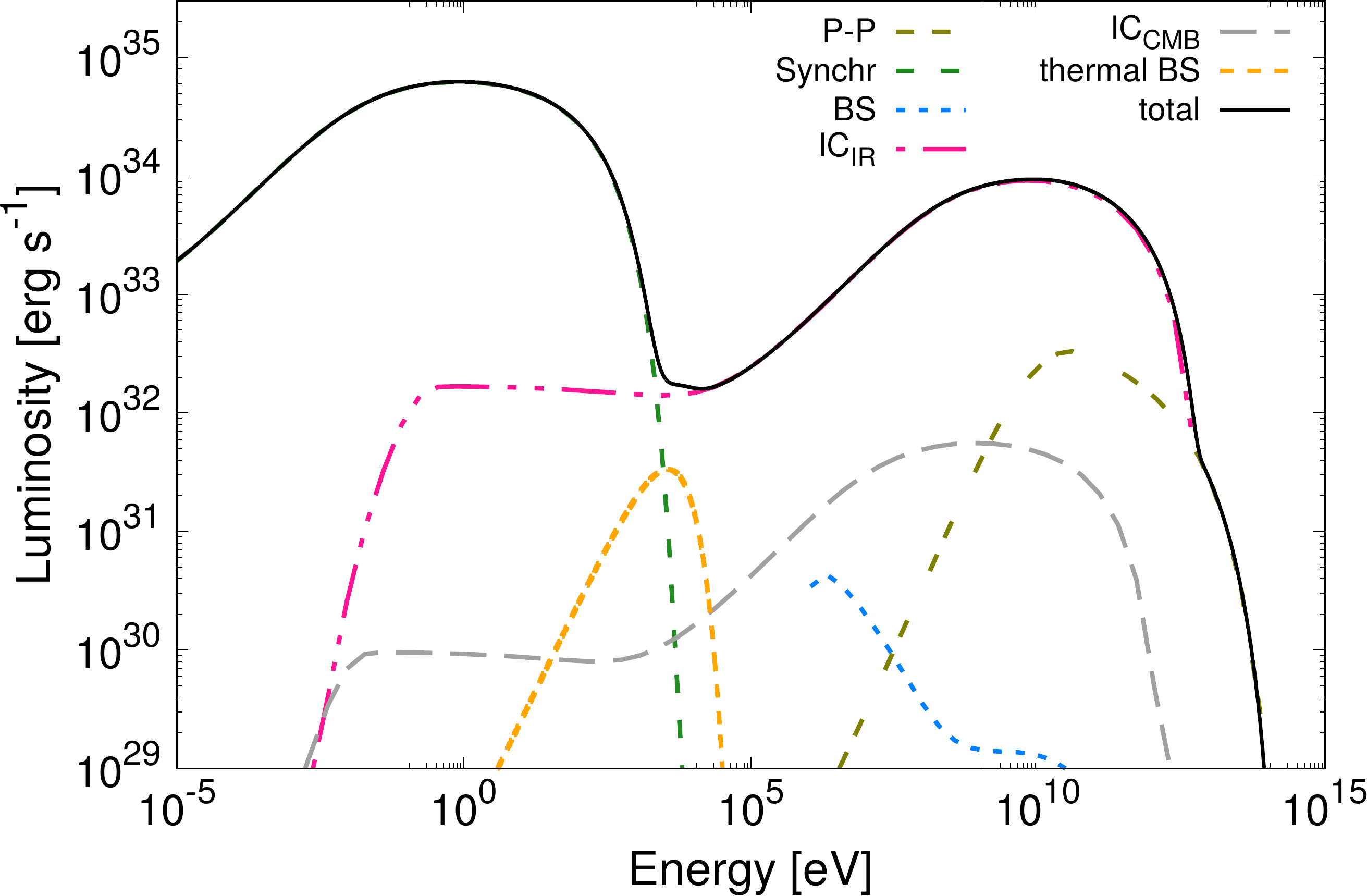}\label{fig:SEDM1a1}}
\subfigure{\includegraphics[trim= 0cm 0cm 0cm 0cm, clip=true, width=.46\textwidth,angle=0]{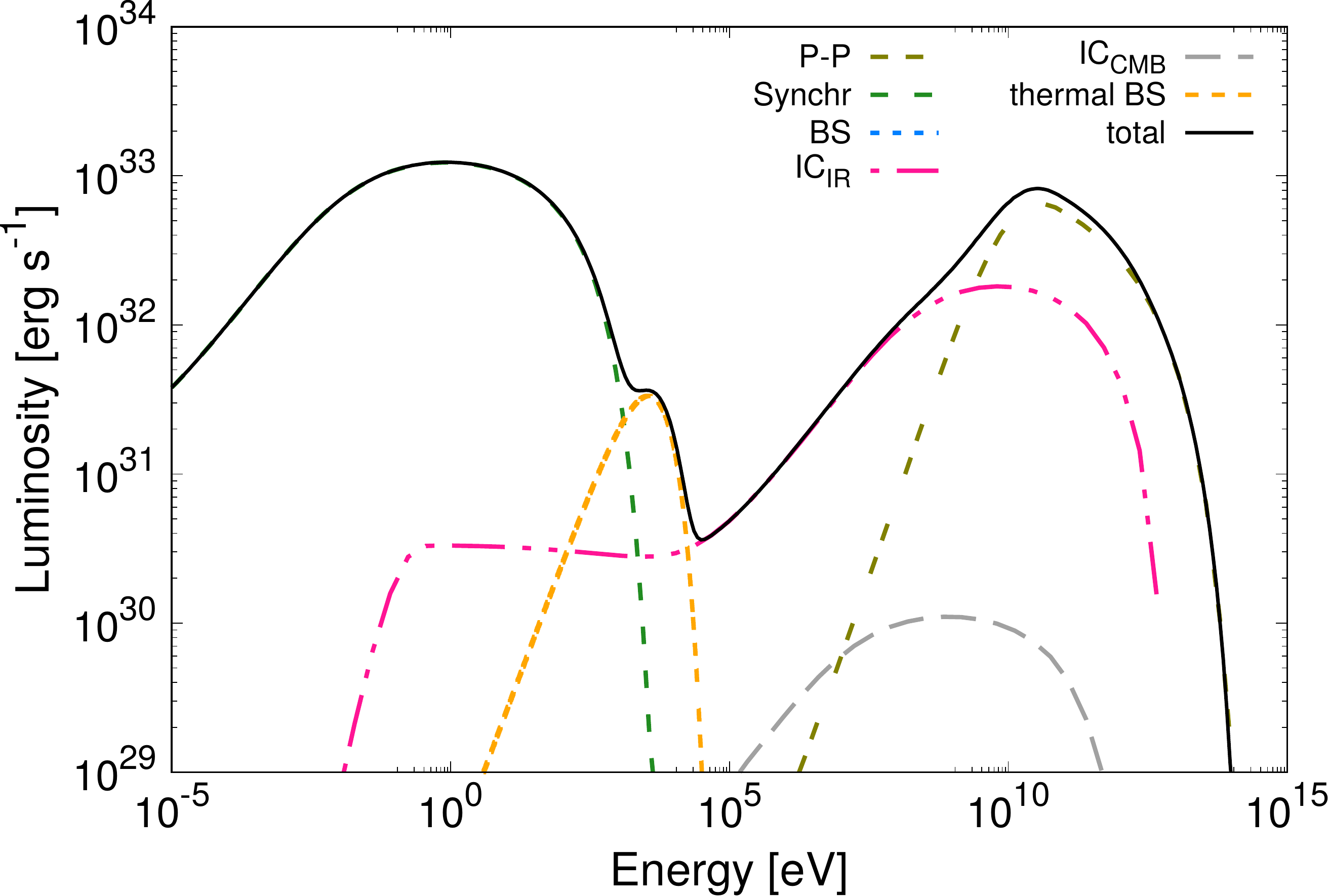}\label{fig:SEDM1a100}}
\caption{Spectral energy distribution for the model M1. The left panel shows the SED obtained with equipartition of energy between accelerated electrons and protons ($L_{p}/L_{e}=1$). The right panel shows the SED assuming $100$ times the energy in electrons to accelerated protons ($L_{p}/L_{e}=100$).} 
\label{fig:SEDM1}
\end{figure*}

\begin{figure*} 
\hspace{-0.15cm}\subfigure{\includegraphics[trim= 0cm 0cm 0cm 0cm, clip=true, width=0.453\textwidth,angle=0]{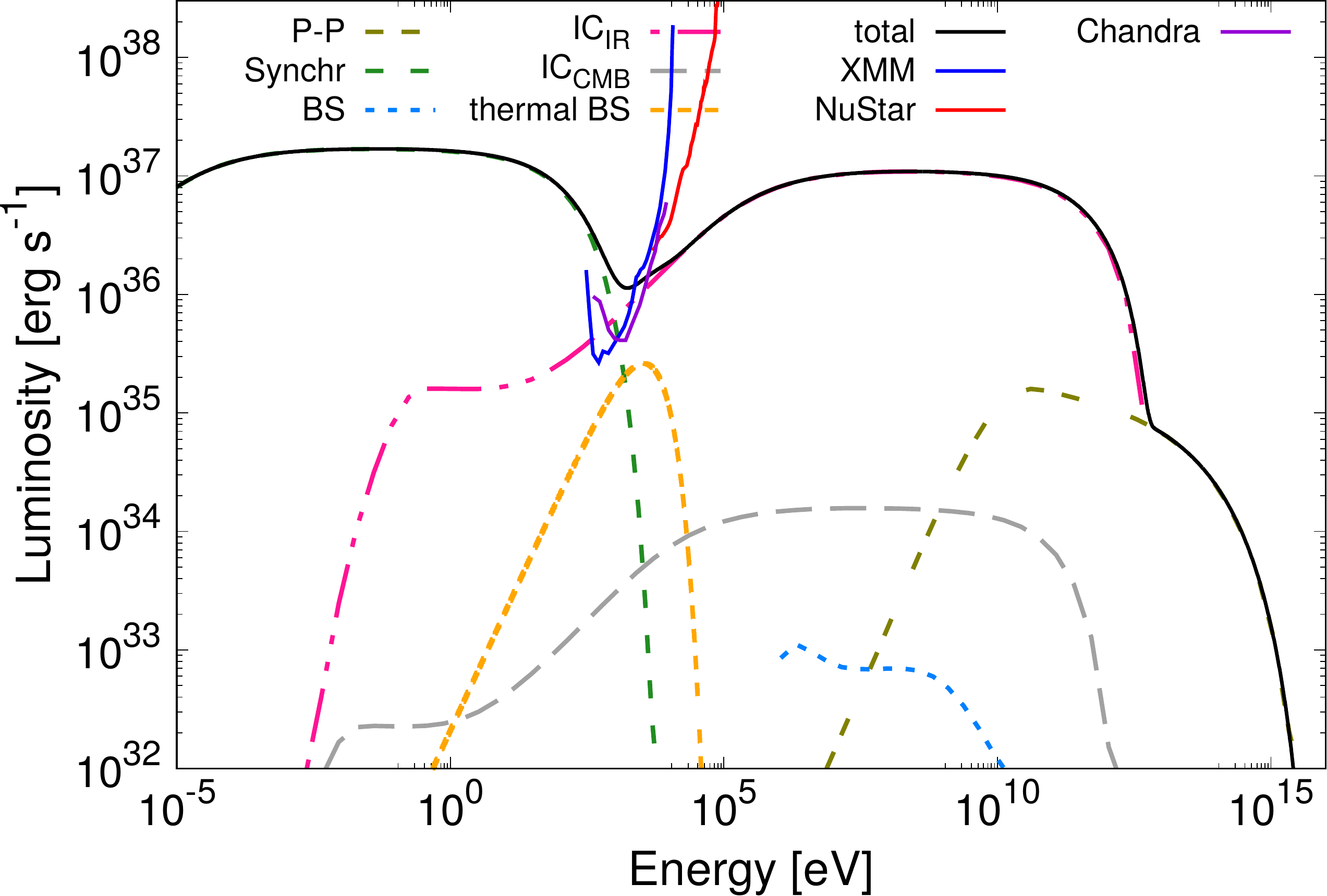}\label{fig:SEDM2a1}}\hspace{0.25cm}
\subfigure{\includegraphics[trim= 0cm 0cm 0cm 0cm, clip=true, width=0.453\textwidth,angle=0]{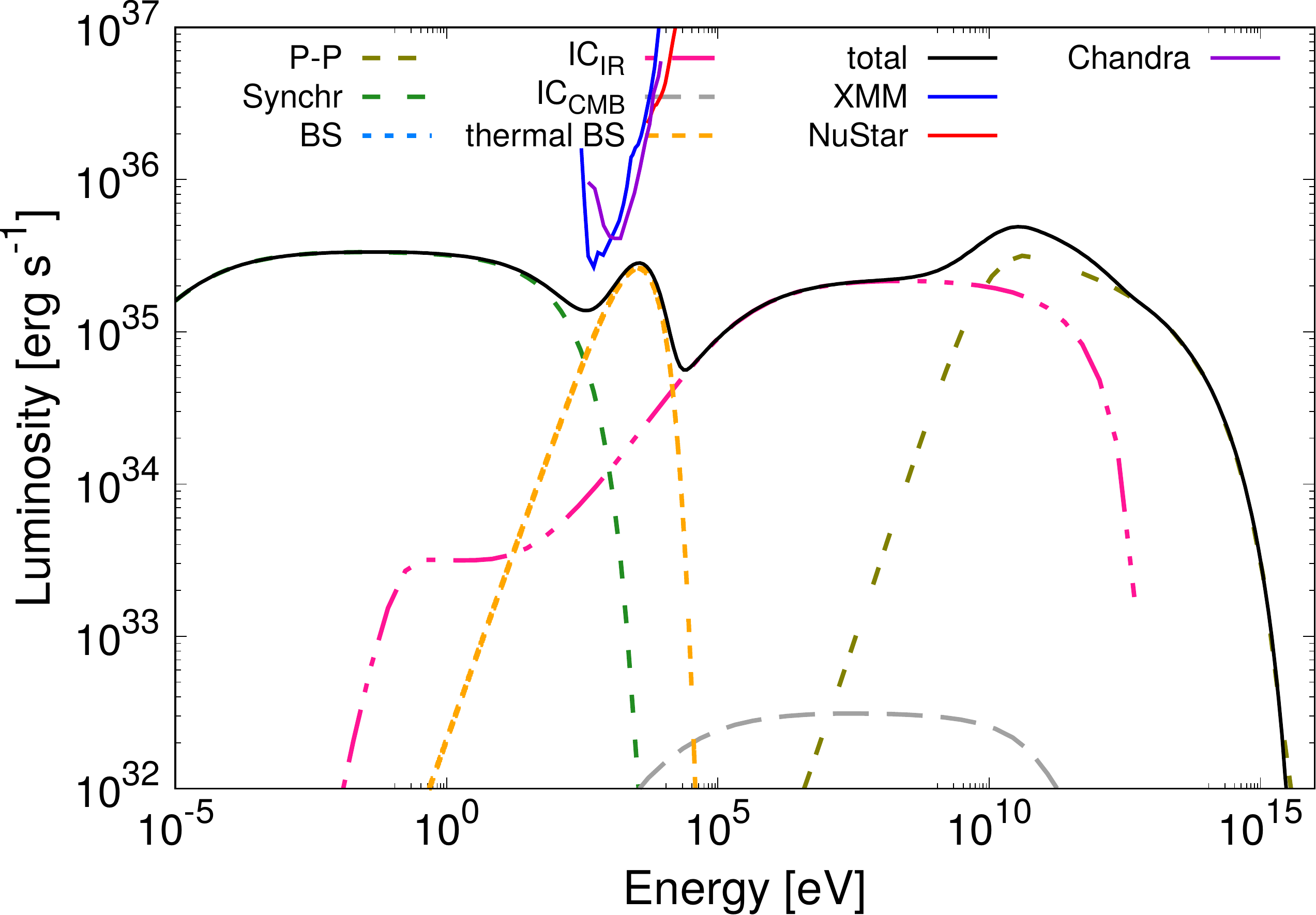}\label{fig:SEDM2a100}}
\caption{Spectral energy distribution for the model M2. The left panel shows the SED obtained with equipartition of energy between accelerated electrons and protons ($L_{p}/L_{e}=1$). The right panel shows the SED assuming $100$ times the energy in electrons to accelerated protons ($L_{p}/L_{e}=100$). The solid lines represent the sensitivity curves of \textit{Chandra}, \textit{XMM-Newton} and \textit{NuStar} for an object at the distance of NGC 253 or M82.} 
\label{fig:SEDM2}
\end{figure*}

In order to estimate the radiation from the particles accelerated in the bowshock, we first calculate the distribution in energy of both electrons and protons solving the transport equation:
\begin{equation}
  \frac{\partial N_{e,p}(E, t)}{\partial t}+\frac{\partial[b(E) N_{e,p}(E,t)]}{\partial E}+ \frac{N_{e,p}(E, t)}{t_{\textrm{esc}}}=Q(E). \label{Teq}
\end{equation}
Here $Q(E)$ is the injection term (a power law with index close to $-2$), $b(E)=\dot{E}$ represents the sum of all the different radiative losses, and $t_{\textrm{esc}}$ is the escape time, which is contributed by the convection and diffusion timescales defined before. The kinetic power of the adiabatic shock is \mbox{$L_{\rm kin}\approx (1/2)\,n_{\rm w}\,m_p\,v_{\rm sw}^{3}\,A_{\rm shock}$}, where $A_{\rm shock}$ is the surface area of the shock \citep{lehnert1999}. We assume that the curvature of the bowshock is negligible along a quarter of the surface area of the sphere of radius $R_{\rm c}+x$ centered in the cloud, therefore \mbox{$A_{\rm shock}=2\,\pi\,(R_{\rm c}+x)^{2}$}. We get \mbox{$L^{\rm M1}_{\rm kin}\sim4.83\times10^{37}$ erg s$^{-1}$} for model M1 and \mbox{$L^{\rm M2}_{\rm kin}\sim1.55\times10^{40}$ erg s$^{-1}$} for model M2. We compute two cases: one where 10\% of this power goes to relativistic particles and is equally distributed among protons and electrons ($L_{p}/L_{e}=1$). The other case is where the power goes more efficiently to protons, with a ratio of proton to electron power of 100 ($L_{p}/L_{e}=100$). The resulting spectral energy distributions (SED) are shown in Figs. \ref{fig:SEDM1} and \ref{fig:SEDM2}. We also include the thermal Bremsstrahlung from the shocked wind material at the bowshock. The IC upscattering of this radiation field is not calculated because its energy density is much smaller than the energy density of the IR or CMB photons. The thermal radiation from the cloud is neglected considering that the initial temperature is at most of \mbox{$10^{4}$ K} \citep{marcolini2005} and the shock propagating through the cloud is too slow to heat it up.

We find that in the case of equal share of energy between electrons and protons, IC dominates at high energies. If hadrons are favored, as in Galactic cosmic rays, then the $pp\rightarrow pp + \pi^0$ channel produces the bulk of high-energy radiation. For small clouds, the absolute maximum of the luminosity predicted by model M1 is \mbox{$\sim 7\times10^{34}$ erg s$^{-1}$} and it is reached at optical wavelengths, whereas the maximum in the $\gamma$-ray band has a value of \mbox{$\sim 10^{34}$ erg s$^{-1}$}. On the other hand, Fig. \ref{fig:SEDM2} shows that big clouds produce higher luminosities. The maximum value, $\sim 10^{37}$ erg s$^{-1}$ is achieved between radio and optical wavelengths, as well as in hard X-rays and soft $\gamma$-rays. We will discuss the detection possibilities in the following section.

\section{Discussion}\label{sec:discu}

The luminosities caused by a single M1-like event are too low to be detected by current instrumental facilities, even for NCG 253 or M82, the nearest starburst galaxies. Although clumps inside superwinds have been observed and cloud-wind interaction models explain successfully the measured soft X-rays, as well as the optical emission and absorption lines, the total number of clouds inside superwinds is not well known. The velocities inferred from optical and ultraviolet lines cannot be associated to single clouds. Furthermore, the simulations indicate that multiple small embedded clouds are expected, but the constraints imposed by cell resolution problems do not allow to obtain reliable estimates of the number of objects \citep{suchkov1994,strickland2000,cooper2009}. The interaction of the denser material that forms the clouds with the more diffuse hot wind gas of the starburst is thought to give rise to \ion{O}{IV} emission and absorption in the far ultraviolet (FUV). The  \ion{O}{IV} absorption lines are then a good tracer of the embedded clouds. \citet{marcolini2005} developed a series of simulations and concluded that to achieve the observational \ion{O}{IV} absorption densities, $\sim 10-30$ clouds with radii between $15$ and $45$ pc in the line of sight are needed. If we assume a typical superwind bubble radius of $5$ kpc and clouds homogeneously distributed, we can roughly estimate that at least $9$ clouds per kpc$^{3}$ should exist. This means, altogether, $\sim 5000$ clouds. As a consequence, it would be possible to detect the integrated luminosity produced by a cluster of $\sim 100$ M1 clouds (\mbox{$R_{\rm c}=5$ pc}) at soft X-ray energies. We also calculate the SED for a cloud under the conditions of \citet{marcolini2005} T1HP model\footnote{We do not include those plots here because the shape of the SED is quite similar to that of our model M2, just with the luminosities 2 orders of magnitude smaller.}. A bunch of $10$ of these \mbox{$15$-pc} clouds could also produce detectable soft X-ray radiation. Regions with multiple small clouds could be created by the fragmentation of larger clouds.

On the other hand, it is not expected to have many M2-like events. As we mentioned in the previous section, the SED for model M2 reaches its maximum at $\gamma$-ray energies. The typical $\gamma$-luminosities for nearby starbursts vary from $6\times10^{39}$ to $1.5\times10^{40}$ erg s$^{-1}$ \citep{ackermann2012,Fermi3rdCatalogue} and the current resolution of the $\gamma$-ray observatories makes it impossible to distinguish the radiation from a single M2 cloud from the total $\gamma$-emission. One of these events could contribute up to $\sim 1 \%$ to the total observed $\gamma$-radiation. In the same way, if we assume that we have $5000$ M1 events, the contribution of the sum of all these events is less than $\sim 1 \%$ to the total luminosity of a starburst galaxy. Assuming $5000$ clouds of \mbox{$15$-pc} radius (\citet{marcolini2005} model), the $\gamma$-ray flux could increase up to $\sim 10 \%$. Resolving a single M2-like event will be possible in the future using the forthcoming CTA observatory by an observation of more than \mbox{$50$ hours} in the case of nearby starburst galaxies (for the sensitivity of CTA see \cite{Hassanetal2017}). 

The soft X-ray radiation produced in the M2 scenario with electron-proton equipartition is large enough to be detected by \textit{XMM-Newton} and \textit{Chandra} (see Fig. \ref{fig:SEDM2}, left panel) in a galaxy at the distance of NGC 253 or M82 ($\sim 3$ Mpc, see \citealt{Dalcanton2009}). If the acceleration of hadrons were preferred, the bowshock thermal Bremsstrahlung could also allow the detection with these satellites. Some of the point-like sources already observed in NGC 253 \citep{strickland2002,bauer2008,wik2014} could actually be associated with the radiation from bowshocks around large clouds or cluster of smaller clouds. Other candidates are X-ray binaries expelled from the galactic disk. The spectra provided in our work, which are quite different from those of accreting binaries either in the low-hard or the high-soft states,  can be used as templates to investigate the nature of individual sources in nearby starbursts such as NGC253 and M82.

The diffuse X-ray halo emission of NGC 253 can be well fitted by two thermal plasma models or a thermal plasma plus a power law. This ambiguity has been discussed by several authors \citep[see for e.g,][and references therein]{strickland2002,bauer2008}, but not solved yet. Our predictions show that small unresolved clouds could contribute to the non-thermal component of this diffuse emission.

\section{Summary and conclusions}
\label{sec:concl}

In this work we have analyzed the acceleration of particles at bowshocks generated around clouds embedded in the hot superwind of starburst galaxies. During recent years starbursts were pointed out as good candidates for sources of ultra high-energy cosmic rays, on the grounds of their astrophysical conditions. Starburst episodes release not only abundant nuclei heavier than protons, but also a great amount of energy into the galactic halo. Some of these particles can become relativistic in large scale shocks. This is supported by the observed high-energy emission associated with nearby galaxies.

We presented the results of two models, whose parameters were chosen to agree with previous simulations \citep{cooper2009,sparre2019}. We assumed local magnetic field amplification and diffusive shock acceleration in bowshocks embedded in the superwind. Although the set of parameters adopted in our individual models are on the extremes of the full range of physical possibilities, namely small and large clouds, we have sensibly extrapolated the results towards the effects of several large clouds, which are expected to dominate the non-thermal emission.

We found that the losses suffered by the relativistic hadrons are dominated by non-radiative processes. Since superwinds seem not to be too dense, those particle could propagate and be reaccelerated in other sites, reaching even higher energies. This possibility will be explored in a future work.

On the other hand, the high-energy electrons cool down locally due to synchrotron and IC scattering with the IR photon field originated in the starburst region. If the energy injected into hadrons does not exceed excessively the energy that goes to electrons, the radiation produced by a bowshock around a large cloud could be detected at X-ray energies by \textit{XMM-Newton} or \textit{Chandra} satellites. This astrophysical situation could actually correspond to some of the point-like X-ray sources observed in \mbox{NGC 253} and \mbox{M 82}.

 The $\gamma$-emission caused by a single large cloud might be detected by CTA in the future, according to our calculations. In the case of small clouds, we conclude that their radiation can only contribute to the diffuse X-ray emission observed in the superwind. Nevertheless, the number of small clouds is expected to be quite large and multiple simultaneous events are expected from the fragmentation of bigger clouds. Therefore clumps of tens or hundreds of clouds with radii of \mbox{$5-15$ pc} could be detected in the X-ray band above the diffuse background.


\section*{Acknowledgements}
We would like to thank the anonymous reviewer for her/his suggestions and comments. GER is very grateful to the IKP at KIT where part of this research was done. This work was supported by the Helmholtz Association through a Helmholtz International Fellow Award to GER. Additional support was provided by the Argentine agencies CONICET (PIP 2014-00338), ANPCyT (PICT 2017-2865) and the Spanish Ministerio de Econom\'{i}a y Competitividad (MINECO/FEDER, UE) under grant  AYA2016-76012-C3-1-P and PID2019-105510GB-C31. 

\section*{Data availability}
The calculations presented in this work were performed using a private code developed and owned by the corresponding author, please contact her for any request/question about. Data appearing in the figures are available upon request. Nevertheless, the results can be reproduced with any code capable of solving the equations indicated in the text using the parameters displayed in the tables.




\bibliographystyle{mnras}

\bibliography{myrefs8}   





\bsp	
\label{lastpage}
\end{document}